\documentstyle[12pt,prl,aps,epsfig,citesort,psfig19,floats]{revtex}
\renewcommand {\baselinestretch}{1.0}

\def\x{{\mbox{\boldmath$x$}}}
\def\u{{\mbox{\boldmath$u$}}}
\def\r{{\mbox{\boldmath$r$}}}

\def\k{{\mbox{\boldmath$k$}}}

\def\e{{\mbox{\boldmath$e$}}}

\def\dz{{\delta\zeta}}
\def\dx{{\delta\xi}}
\def\la{{\langle}}
\def\ra{{\rangle}}

\def\eps{{\epsilon}}

\def\rel{{Re_\lambda}}

\def\begineq{\begin{equation}}
\def\endeq{\end{equation}}
\def\be{\begin{equation}}
\def\ee{\end{equation}}

\begin{document}
\bibliographystyle{prsty}

\title{
Different intermittency for longitudinal and transversal 
turbulent fluctuations}
\author{Siegfried Grossmann,
Detlef Lohse, and
Achim Reeh}
\address{
Fachbereich Physik der Universit\"at Marburg,\\
Renthof 6, D-35032 Marburg, Germany}

\date{\today}

\maketitle
\begin{abstract}
Scaling exponents of the longitudinal and transversal velocity structure
functions
in numerical Navier-Stokes turbulence simulations with
 Taylor-Reynolds numbers up to  $\rel
= 110$
are determined 
by the extended self similarity method.
We find significant differences in the degree of intermittency:
For the sixth moments the scaling
corrections to the classical Kolmogorov expectations
are $\delta\xi_6^L= -0.21 \pm 0.01$ and $\dx_6^T= -0.43 \pm 0.01$, 
respectively, independent of $\rel$.
Also the generalized extended self similarity exponents
$\rho_{p,q} = \dx_p/\dx_q$ 
differ significantly
for the longitudinal and transversal structure functions.
Within the She-Leveque model this means that longitudinal and transversal
fluctuations obey different types of hierarchies of the moments. 
Moreover, the She-Leveque model hierarchy parameters $\beta^L $ and $\beta^T$
show small but significant dependences 
on the order of the moment. 
\end{abstract}

\newpage


\section{Introduction}
One of the central issues in turbulence theory has always been whether the
velocity structure functions deviate from Kolmogorov's classical expectation
\cite{my75,fri95}. For many years the community focused on the {\it
longitudinal} structure function 
\begin{equation}
  D_{p}^L (r) =
  \langle [
(\u (\x +\r) - \u (\x )) \cdot \e_r^L
  ]^p \rangle
  =\langle (v^L (r) )^p\rangle
\label{eq1}
\end{equation}
whose inertial subrange (ISR) scaling exponents we define as $\zeta_p^L$.
(Here, $\e_r^L$ is the unit vector in $\r$ direction.) The reason for this
was that in many experiments Taylor's frozen flow hypothesis
\cite{tay38,my75,fri95}
had to be employed and therefore, the transversal structure functions
\begin{equation}
  D_{p}^T (r) =
  \langle [
(\u (\x +\r) - \u (\x )) \cdot \e_r^T
  ]^p \rangle 
  =\langle (v^T (r) )^p\rangle ,
\label{eq2}
\end{equation}
$\e_r^T$ being a unit vector perpendicular to $\r$, could not be obtained.
We denote the scaling exponents of $D_p^T(r)$ as $\zeta_p^T$.

A priori, there is no reason to expect $\zeta_p^L =\zeta_p^T$ for general $p$.
The PDF of the {\it longitudinal}
 velocity difference $v^L(r)$ is skewed because
information from $\x$ to $\x + \r$ is conveyed by the velocity difference
itself, and odd moments of $v^L(r)$ thus do not vanish, $D_p^L\ne 0$.
In particular,
for homogeneous and isotropic turbulence
the third order longitudinal structure function is connected to
the second order one  by
the Howard-v.\ Karman-Kolmogorov
 structure equation \cite{my75,fri95},
\be
D_3^L (r) =- {4\over 5} 
\eps r + 6\nu {d\over dr} D_2^L (r).
\label{eq3}
\ee
Here, $\eps$ is the mean energy dissipation rate 
and $\nu$ is the kinematic viscosity. The PDF of the {\it transversal }
velocity difference $v^T(r)$, on the other hand, is symmetric and consequently
all odd order moments vanish, $D_p^T(r) =0$, $p$ odd.
Only
the second order 
longitudinal and transversal structure functions 
 are expected to
 scale the same way
\cite{nou97,my75,oul96}
  in the large Reynolds number limit
 since
 for isotropic flow incompressibility implies \cite{my75}
\be
D_2^T (r) = D_2^L(r) + {r\over 2} {d\over dr} D_2^L (r).
\label{eq4}
\ee

Recently, multi-probe and optical techniques made the transversal structure
functions experimentally accessible
\cite{sad94,her95,her95,her97,cam96,cam97,nou97}. In addition,
also recent numerical simulations
of decaying turbulence
\cite{bor97}
have focused on the difference in scaling of
$D_p^L(r)$ and $D_p^T(r)$.

The results of all this work seem to be contradictory.
While all authors agree that (i) both $\zeta_p^L$ and $\zeta_p^T$
show significant deviations from the classical expectation $p/3$,
and that (ii) $\zeta_p^L$ is well fitted by the She-Leveque model
(``SL model'')
\cite{she94}, saying that
\begineq
\zeta_p = {p\over 3} - C_0 \left( {p\over 3} (1-\beta^3 ) - (1-\beta^p )
\right) 
\label{sl}
\endeq
with $C_0 =2$ and $\beta = (2/3)^{1/3}$,
they disagree on whether
$\zeta_p^L = \zeta_p^T$ or
$\zeta_p^L \ne \zeta_p^T$.

Van der Water's group \cite{her95,her97} finds that
in shear flow the (moduli of the)
transversal intermittency corrections
$\dz_p^T =\zeta_p^T - p/3$ are
significantly larger than the longitudinal ones
$\dz_p^L =\zeta_p^L - p/3$, $p>3$. The same is found for
jet flow turbulence by
Camussi and Benzi \cite{cam97} and for
decaying
(numerical) turbulence by Boratav and Pelz \cite{bor97}. 
On the other hand, the recent experiments by
Camussi et al.\
\cite{cam96} and Kahalerras et al.\
\cite{kah96} did not give significant deviations between
$\dz_p^L$ and $\dz_p^T$ and experiments by
Noullez et al.\ \cite{nou97} found $\dz_p^T$ comparable to 
the $\dz_p^L$ found in other experiments. 
For a simple quantification of the intermittency corrections it is common to
give $\dz_6$. We do so in table \ref{tab_exp}
 for the addressed experiments and
simulations.

 \begin{table}[htp]
 \begin{center}
 \begin{tabular}{|c|c|c|c|c|c|}
 \hline
          ref.\ 
       &  flow 
       & $\rel$
       & $-\dx_6^L$
       & $-\dx_6^T$
       &  remark
 \\
\hline
        van der Water et al.\ \cite{her97}   
       & shear flow
       & up to $600$
       & $0.18 - 0.20$
       & $0.27-0.31$
       & normalized
\\
         Camussi and Benzi \cite{cam97}   
       & jet flow
       & $250$
       & $0.25$
       & $0.38$
       & ESS 
\\
        Camussi et al.\ \cite{cam96}   
       & wind tunnel flow
       & $37$
       & $0.19 \pm 0.03$
       & $0.22\pm 0.03$
       & ESS
\\
        Boratav and Pelz \cite{bor97}   
       & decaying numerical flow
       & $\sim 100$
       & $0.23$
       & $0.43$
       & ESS
\\
        Noullez et al \cite{nou97}   
       & jet flow
       & up to $600$
       & $-$
       & $0.25\pm 0.10$
       & ESS
\\
        this work 
       & forced numerical  flow
       & $110$ 
       & $0.21\pm 0.01$
       & $0.43\pm 0.01$
       &  ESS
\\
        this work 
       & forced numerical  flow
       & $70$ 
       & $0.22\pm 0.01$
       & $0.43\pm 0.01$
       &  ESS
\\
 \hline
 \end{tabular}
 \end{center}
\caption{{\it
Intermittency corrections
$\dx^L_6$ and $\dx^T_6$ (see text for the definition)
for the longitudinal and transversal
sixth order
structure functions for various numerical and experimental flows.
}}
\label{tab_exp}
 \end{table}

We would like to caution the reader of a too simplistic interpretation
of table 1. The detailed definitions of the $\dz_p^{L,T}$
slightly differ from
experiment to experiment. In all low Reynolds number experiments
\cite{nou97,cam96,cam97,bor97} the scaling exponents could only
be determined 
by employing the extended self similarity (ESS) method introduced by
Benzi et al.\ \cite{ben93b}. In this method the structure functions $D_p(r)$
are plotted against $D_3^*(r)$, where $D_3^*(r)$ is the third order structure
function defined with the {\it modulus} of the velocity difference and is
experimentally found to scale with roughly the same exponent as $D_3(r)$,
which includes the sign of $v(r)$ \cite{ben95}.
The ISR scaling exponent of such a plot is
henceforth denoted as $\xi_p = \zeta_p/\zeta_3^*$.
As in the ISR
$\zeta_3^L = 1$ according to the structure equation (\ref{eq3}), one
expects $\xi_p^L = \zeta_p^L$, if $\zeta_3^{*L} = \zeta^L_3$.
In experiments, however, there are always
small deviations from $\zeta_3^{*L} =1$; therefore, in principle $\zeta_p^L$
 and $\xi_p^L$ could slightly differ, see also ref.\
 \cite{gro97c}.
For the transversal structure function
 $D_3^{*T}(r)$ there is no known
 relation as equation (\ref{eq3}). Indeed,
 van der Water's group finds $\zeta_3^{*T}=1.08$ \cite{her97} and they
  give {\it normalized}
 scaling exponents $ \zeta_p /\zeta_3$ which we also call  $\xi_p$.
To date, there is no strict theoretical argument why ESS works so
well. Note that for the present numerical simulation it does {\it not}
work for odd order structure functions, taken {\it without} the modulus
\cite{gro97c}.

What is the
origin of the differences between the results reported in table 1? One may think
that it is the different geometry of the flows which causes the differences, in
particular, a different strength of the
local shear and of the anisotropy in the flow.
Indeed, the shear in the flow of ref.\ \cite{her97} is considerable and it is
known that shear destroys ESS \cite{sto93b,ben96b}. On the other hand,
at least
$\dx_6^L$ was found to be remarkably independent 
of different flow geometries \cite{arn96}. Also, the 
numerical flow of ref.\
\cite{bor97} which clearly shows $\dx_6^L\ne \dx_6^T$ is
highly isotropic. But it is decaying which in ref.\ \cite{bor97b} is speculated
to be a possible origin of the observed \cite{bor97}
discrepancy between longitudinal and
transversal intermittency corrections.

In this paper we set out to determine the scaling exponents of the longitudinal
and transversal structure functions for forced, statistically stationary
numerical Navier-Stokes turbulence up to $\rel =110$.
Our motivation is to contribute to clarify the contradictionary
picture reflected in table 1. It is of general interest for the
understanding of intermittency whether in Navier-Stokes dynamics
not only the velocity field and derivatives thereof are independently
scaling fields as analyzed in ref.\ \cite{itamar}, but that there are already
two independently scaling velocity fields $v^L (r)$ and $v^T(r)$. Indeed,
we will find 
significant differences for the longitudinal and transversal scaling
corrections, namely, to put the result in a nutshell, 
$\dx_6^L = 0.21\pm 0.01$ and
$\dx_6^T = 0.43 \pm 0.01$ in very good agreement with Boratav's result for
decaying turbulence \cite{bor97}.

The question which immediately comes up is whether this finding originates in
the anisotropy of the flow. Therefore, we carefully analyze the degree of
anisotropy of the numerical flow. We find only limited anisotropy and only on
the very large scale and therefore consider the different scaling of
longitudinal and transversal structure functions the more remarkable.

The second point we examine is whether these different intermittency
corrections $\dx_p^L$ and $\dx_p^T$
correspond to different {\it hierarchies } of the moments. Such hierarchies were
suggested by She and Leveque \cite{she94}
for the r-averaged energy dissipation rate $\epsilon_r$, namely
\be
{\la \eps_r^{p+1}\ra  \over 
\la \eps_r^p \ra} = B_p'
\left({  \la \eps_r^p \ra \over 
\la \eps_r^{p-1}\ra }\right)^{\beta^3}
 (\eps_r^{(\infty )})^{1-\beta^3},
 \label{eps_hier}
 \ee
$B_p'$ constant,
 $\eps_r^{(\infty )} = \lim_{p\to \infty } (\la \eps_r^{p+1} \ra
 / \la \eps_r^p\ra )$; the SL parameter $\beta$ is therefore called
 hierarchy parameter.
Such a hierarchy means that the corresponding probability distribution function
 obeys a log-Poisson
statistics
\cite{she95}.

Ruiz Chavarria et al.\ \cite{rui95} extended the idea of hierarchies
to structure
functions.
Assuming Kolmogorov's refined similarity hypothesis
$D_p (r) \sim \la \eps_r^{p/3} \ra r^{p/3}$ 
\cite{kol62,fri95} the structure
function hierarchy can be derived from
the SL hierarchy
 (\ref{eps_hier})
\cite{rui95,ben96b}
and reads
\be
{D_{p+1} (r)  \over D_p (r) }
= B_p'' \left( {D_p(r) \over D_{p-1}(r) }\right)^\beta
\left( D_{(\infty)} \right)^{1-\beta },
\label{d_hier}
\ee
$B_p''$ constant, $D_{(\infty )}(r) = \lim_{p\to \infty}
(D_{p+1}(r)/D_p(r)) =
(r\eps_r^{(\infty )} )^{1/3}$.

We will calculate the hierarchy parameter $\beta$ both for the longitudinal and
transversal structure functions, very carefully considering the systematic and
statistical errors. First, we find significant deviations between $\beta^L$ and
$\beta^T$. Second, we find a slight but also significant dependence of the
hierarchy parameters $\beta^L$ and $\beta^T$ on the order of the moment which is
not expected within the SL model.

The paper is organized as follows: In section 2, we define the numerical flow
and carefully
check its isotropy, in section 3 we report on various scaling relations,
employing ESS and the generalized ESS (GESS, \cite{ben96b,ben95});
we also calculate the hierarchy parameters $\beta^L$ and $\beta^T$. 
In section 4 we determine $\dx_p^L$ and $\dx_p^T$ within a reduced wave
vector set approximation of the Navier-Stokes dynamics
\cite{egg91a,gnlo92b,gnlo94a} in which very large $\rel$ can be
achieved.
Conclusions are drawn in section 5.

\begin{figure}[htb]
\setlength{\unitlength}{1.0cm}
\begin{picture}(9,9)
\put(0.5,0.5)
{\psfig{figure=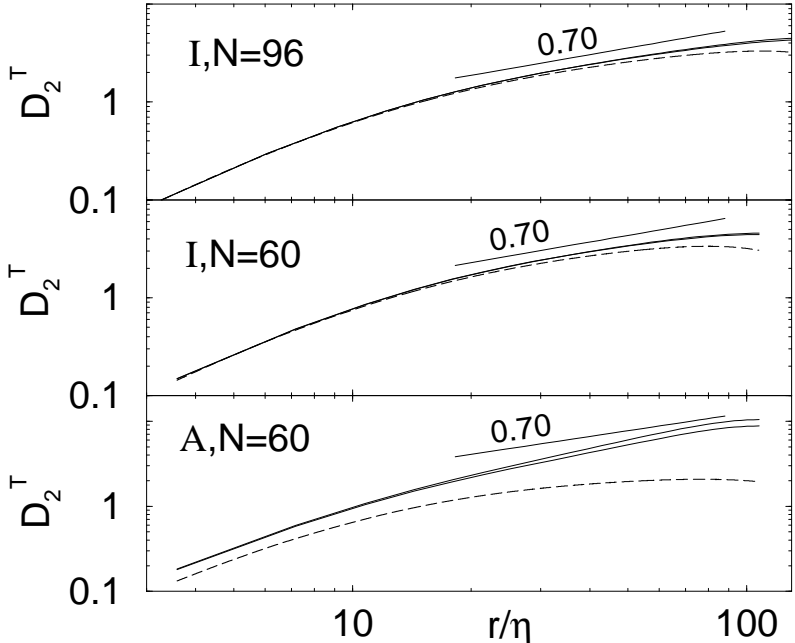,width=10cm,angle=-90}}
\end{picture}
\caption[]{
Second order {\it transversal} structure functions $D_2^T(r)$ for
the isotropic $N=96$, $\nu = 0.006$ simulation ($\rel = 110$, upper),
the isotropic $N=60$, $\nu = 0.009$ simulation ($\rel = 70$, middle),
and the anisotropic $N=60$, $\nu = 0.009$ simulation ($\rel = 70$, bottom).
The solid lines are calculated from
the definition (\ref{eq2}) of $D_2^T(r)$ (for the two directions being
perpendicular to $\r$), the dashed line
is calculated from relation (\ref{eq4}) which holds for perfect isotropy
and homogeneity.
For the anisotropic case 
anisotropy can be seen on {\it all} scales; also the ISR slope
deviates from the expected value $\zeta_2 = 0.70$. 
}
\label{fig_aniso}
\end{figure}

\begin{figure}[htb]
\setlength{\unitlength}{1.0cm}
\begin{picture}(9,9)
\put(0.5,0.5)
{\psfig{figure=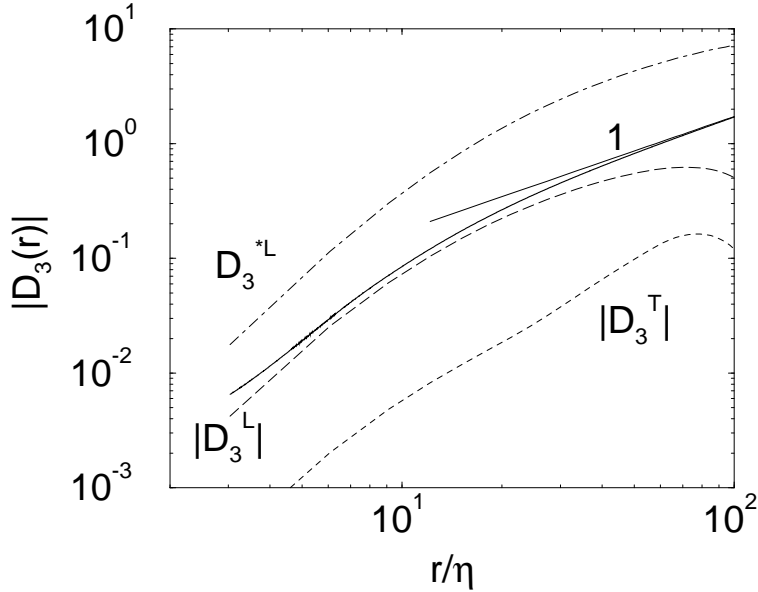,width=10cm,angle=-90}}
\end{picture}
\caption[]{
Third order structure function
$D_3^L(r)$, directly calculated from the numerics
(long dashed) and from Kolmogorov's structure equation
(\ref{eq3}) (solid). Also shown are
 $D_3^{*L}(r)$ (dashed-dotted) and
 $|D_3^{T}(r)|$ (short dashed). 
The data are for the isotropic $N=96$ simulation ($\rel = 110$).
}
\label{fig_d3_vergleich}
\end{figure}

\section{Set up of the flow and check of its isotropy} 
The 3D incompressible
Navier-Stokes equations are numerically solved on a $N^3$
grid with periodic boundary conditions.
Spherical truncation is used to reduce aliasing. 
For the isotropic flow simulation (denoted by ``I'')
we force the system on the largest scale
(wavevectors
$\k = (0,0,\pm 1)/L$,
$\k = (0,\pm 1,\pm 1)/L$,
$\k = (\pm 1,\pm 1,\pm 1)/L$, 
and permutations thereof)
with a forcing term as e.g.\ described in ref.\
\cite{gnlo94a}.
Units are fixed by picking the length scale $L=1$ and the
average energy input rate (= the energy dissipation rate) $\eps=1$.
The Taylor-Reynolds number is defined as
 $\rel= u_{1,rms}\lambda /\nu$,
 where $\lambda = u_{1,rms}/ (\partial_1 u_1)_{rms}$ is the Taylor
 length and 
 $\nu$  the viscosity.
Most of our results refer to
$N=96$ and $\nu =0.006$, corresponding to 
a resolution of scales  $r \ge 2\pi L/N \approx 3\eta$
and $\rel = 110$. 
Time integrations
of about 
140 large eddy turnover times are performed.
Averages are taken over space and
time.
To check the Reynolds number dependence we also did 
an isotropic $N=60$, $\nu = 0.009$ simulation (240 large eddy turnovers)
which has $\rel = 70$.
For a less isotropic flow simulation (denoted by ``A'')
we only force one mode
$\k=(0,0,1)/L$. 
This simulation is done for 
$N=60$, $\nu=0.009$, for about 210 large eddy
turnovers; it has $\rel = 70$, too.

\begin{figure}[htb]
\setlength{\unitlength}{1.0cm}
\begin{picture}(9,9)
\put(0.5,0.5)
{\psfig{figure=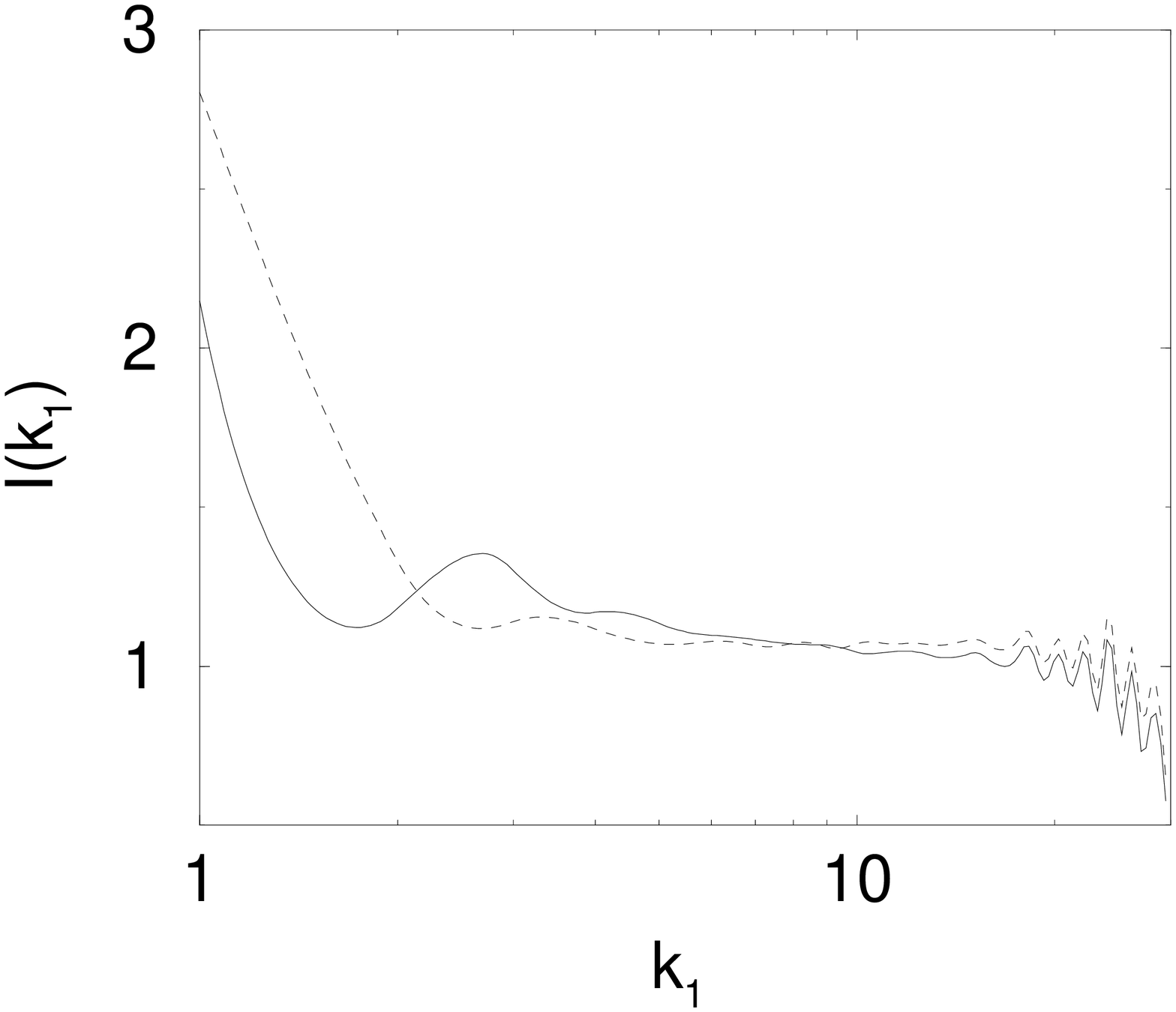,width=10cm,angle=-90}}
\end{picture}
\caption[]{
Isotropy coefficient $I(k_1)$ eq.\
 (\ref{iso_coeff}) for the simulations ``I'' (solid)
and ``A'' (dashed) ($N=60$). 
}
\label{fig_iso_coeff}
\end{figure}

We checked the isotropy of the flow in several ways:
\begin{itemize}
\item
We calculated the structure functions for different space directions
and compared them among each other.
For the simulation ``I'' good agreement is found, for ``A'' one space
direction is distinguished as expected
 from the type of forcing, see figure
\ref{fig_aniso}.
Moreover, for the isotropic simulation we find less than $5\%$ deviations
between
$\la u_1^2\ra$, 
$\la u_2^2\ra$, and
$\la u_3^2\ra$. Note that
$D_2^L(r=\pi)$ and 
$D_2^T(r=\pi)$ (We used $r=\pi$ as the largest 
possible space separation in numerical flow with periodic boundary
conditions.) do {\it not} equal $2\la u_i^2\ra$ ($i=1,2$, or $3$)
as expected for experimental, isotropic flow at $r=\infty$.
 We find
deviations  up to $25\%$ which means that the velocities are still
correlated at the space distance of $r=\pi$. For the longitudinal
velocities we find a positive correlation of about $25\%$, for the
transversal velocities we find a negative correlation of 
about $15\%$. Geometrically, this means that there is a large scale eddy 
with diameter $\sim \pi$. --  We can not
fully exclude that the results
on scaling exponents we will report on
are influenced by the flow geometry (periodic boundary conditions).
They might be different for different geometries (e.g., those in
experimental flows). 
\item
We checked relation (\ref{eq4}) which only holds for isotropy \cite{my75}.
For ``I'' there are
only large scale deviations, for ``A'' deviations show up down to small
scales, see figure \ref{fig_aniso}.
\item
We checked the relation (\ref{eq3}),
see figure \ref{fig_d3_vergleich}. It holds for isotropic
flow.
The agreement is reasonable. However,
there still is no developed inertial subrange due to the low $\rel$.
The curve looks very similar to the experimental curve
for comparable $\rel$, cf.\ fig.\ 2 of ref.\ \cite{her97}.
In particular, also the experimental curves bend down for large $r$. 
The reason for this of course is that at large scales
the fluctuations are Gaussian and odd order moments vanish.
We ascribe the deviations in the viscous subrange (VSR) 
to the lack of perfect convergence of odd moments.
This difference remained even for as long averaging times as 140
large eddy turnovers. 
Also the relation
$D_3^T(r) =0$ is not yet fulfilled for this low $\rel$, though the modulus
of $D_3^T(r) $ is
 more than one decade smaller than the
modulus of $D_3^L(r)$  for all scales, see figure \ref{fig_d3_vergleich}.
\item
For perfect isotropy, the mean
energy dissipation rate $\eps$ can be calculated
from {\it any} component of the strain tensor $\partial_i u_j$, e.g.,
\be
\eps =
15\nu \la (\partial_1 u_1 )^2 \ra =
{15\over 2}\nu \la (\partial_2 u_1 )^2 \ra .
\label{eps_ten}
\ee
For the isotropic flow, these relations hold very well, see table
\ref{tab_aniso}, for the anisotropic one there are deviations
up to $15\%$. 
\item
For an isotropic flow, the isotropy coefficient \cite{jim93}
\be
I(k_1) = { E_{11} (k_1) - k_1 \partial E_{11} (k_1) /\partial k_1
\over 2 E_{22} (k_1)}
\label{iso_coeff}
\ee
which compares the longitudinal and transversal energy spectra
$E_{11} (k_1)$ and 
$E_{22} (k_1)$ should become 1 \cite{bat53}. $I(k_1)$ is shown in figure
\ref{fig_iso_coeff}.
Indeed, in general $I(k_1)$ is closer to 1 for ``I'' than
it is for ``A''.
We do not quite understand 
the bump around $k_1 =2.5$ in $I(k_1 )$ in the isotropic simulation.
We tend to ascribe it 
to the forcing of the modes $(\pm 1 , \pm 1 , \pm 1) /L$.
The wiggles for very large $k_1$ are numerical artifacts
because of the derivative in eq.\
(\ref{iso_coeff}). 
\end{itemize}

 \begin{table}[htp]
 \begin{center}
 \begin{tabular}{|c|c|c|c|c|}
 \hline
          \phantom{} 
       &  energy input
       & $\eps$
       & $15\nu \la (\partial_1 u_1 )^2 \ra $
       & ${15\over 2}\nu \la (\partial_2 u_1 )^2 \ra $
 \\
\hline
         I, $N=96$
       & $1$
       & $1.003$
       & $0.984$
       & $1.021$
\\
         I, $N=60$
       & $1$
       & $1.003$
       & $0.984$
       & $0.994$
\\
         A, $N=60$
       & $1$
       & $1.004$
       & $0.931$
       & $0.851$
\\
 \hline
 \end{tabular}
 \end{center}
\caption{{\it
Energy input and dissipation rates for the three numerical simulations.
The good agreement between the energy input and the total energy
dissipation
rate $\eps$ means statistical stationarity. The degree of agreement
between the last two columns with 1 
characterizes the degree of isotropy in the VSR.
}}
\label{tab_aniso}
 \end{table}

\section{Scaling exponents for longitudinal and transversal structure
functions}

\begin{figure}[p]
\setlength{\unitlength}{1.0cm}
\begin{picture}(9,18)
\put(0.5,9.5)
{\psfig{figure=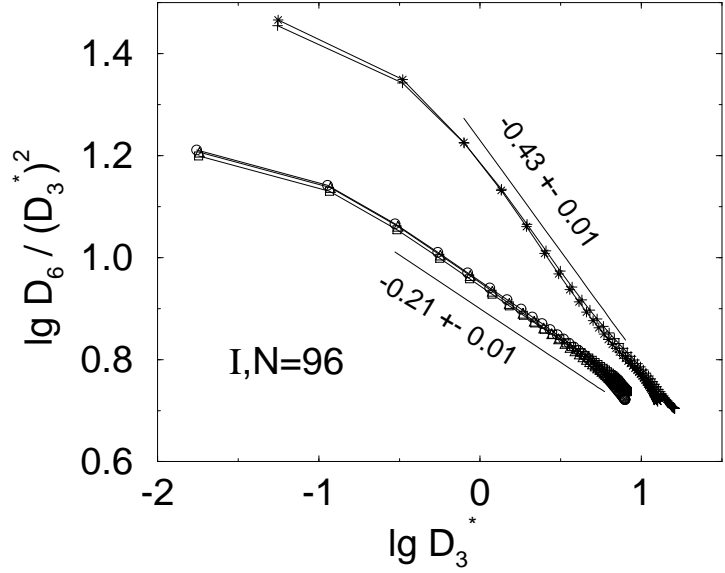,width=10cm,angle=-90}}
\put(0.5,0.5)
{\psfig{figure=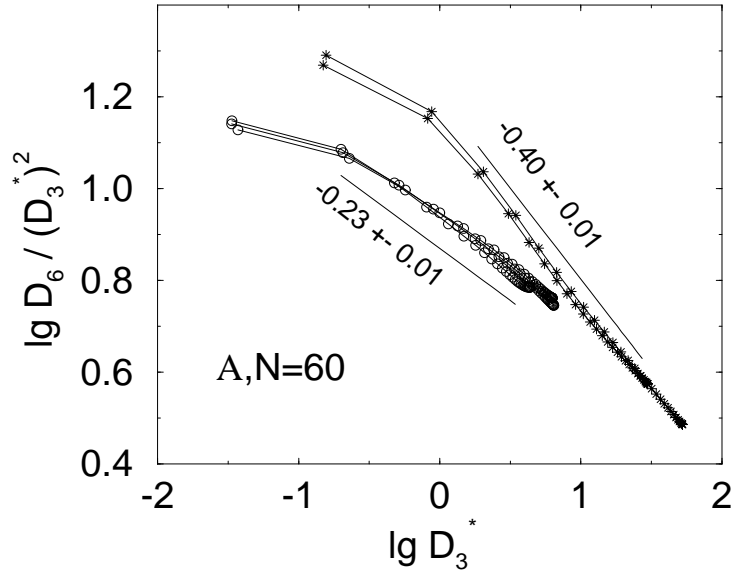,width=10cm,angle=-90}}
\end{picture}
\caption[]{
Compensated ESS type plots for 
$D_6^L/(D_3^{*L})^2$ vs $D_3^{*L}$ (circles) and 
$D_6^T/(D_3^{*T})^2$ vs $D_3^{*T}$ (stars). 
The ISR slopes are respectively $\dx_6^L$ and $\dx_6^T$.
The upper part of the figure refers to the isotropic
simulation with $N=96$, the lower one to the anisotropic one with
$N=60$. 
}
\label{fig_ess6}
\end{figure}

\begin{figure}[htb]
\setlength{\unitlength}{1.0cm}
\begin{picture}(9,9)
\put(0.5,0.5)
{\psfig{figure=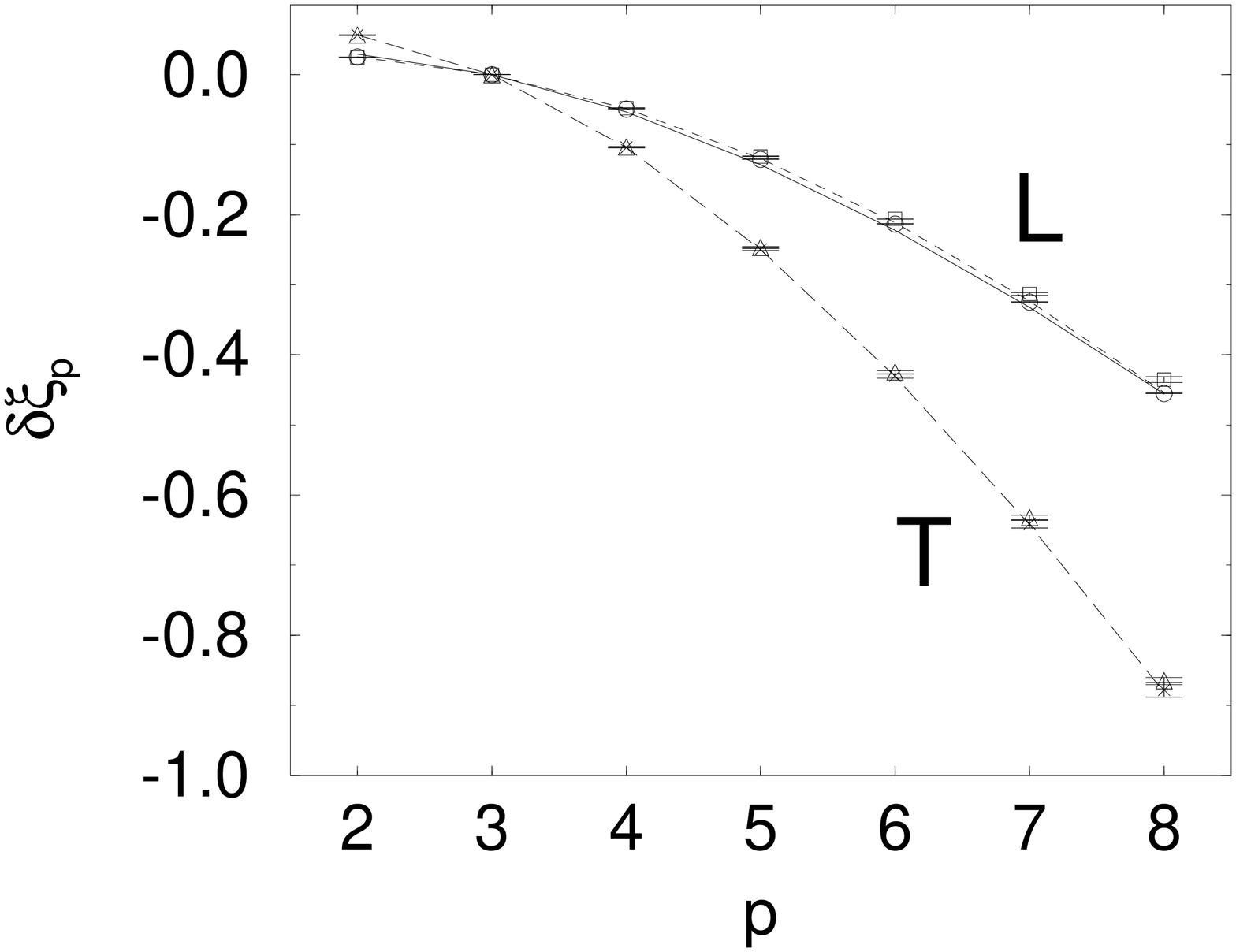,width=10cm,angle=-90}}
\end{picture}
\caption[]{
Intermittency corrections $\dx_p^L $ (circles for $N=96$,
squares $N=60$) and $\dx_p^T$ (crosses $N=96$, triangles $N=60$)
 from the isotropic numerical
simulations.
The dashed lines are 1-parameter fits within the SL-model
where the $\beta$'s have been taken fixed from the averaged fits of the
GESS type plot, i.e.,
$\beta^L = 0.947$ and  
$\beta^T = 0.870$.
The one free fit parameter is thus $C_0$. We obtain the shown
remarkably good 
fits with $C_0^L = 9.3$ for the longitudinal data
(short dashed) and $C_0^T=3.7$ for the transversal data (long dashed).
The standard SL fit \cite{she94}
$\beta = (2/3)^{1/3}$, $C_0=2$ fits the longitudinal corrections
also pretty well, see the solid line.
}
\label{fig_zetasl}
\end{figure}

To determine the degree of intermittency
in the longitudinal and transversal structure functions we employ
a type of ESS
\cite{ben93b,ben94a,ben96b}
by calculating generalized structure functions
\be
G_p(r) = { D_p (r) \over (D_3^{*}(r))^{p/3} }
\label{gfct}
\ee
and plotting them vs $D_3^*$ (``compensated ESS plot'' \cite{gro96,cam96}),
see figure \ref{fig_ess6}.
The intermittency exponents $\dx_6$ (the ISR slopes in figure \ref{fig_ess6})
for the longitudinal and
transversal structure functions are clearly different; the transversal signal
shows considerably more intermittency.
No dependence on $\rel$ is found. The values of $\dx_6^{L,T}$ for
the isotropic $\rel =110$ and $\rel = 70$ simulations are given in table
\ref{tab_exp}. 
Surprisingly, also the 
anisotropic  simulation ``A'' approximately has the same
scaling exponents, namely 
$\dx_6^L=-0.23 \pm 0.01$ and
$\dx_6^T=-0.40 \pm 0.01$, see figure \ref{fig_ess6}. 
Therefore, in what follows we will only focus on the isotropic simulation
``I''.
Our results
for various $\dx_p$ determined as in figure \ref{fig_ess6}
 are summarized in figure
\ref{fig_zetasl}.

It can be seen that the intermittency corrections
$\dx_p^L$ and 
$\dx_p^T$ clearly deviate throughout, i.e., transversal 
velocity fluctuation are much more intermittent than
longitudinal ones. 
Though it has been known for many years that  the transversal 
velocity {\it gradient} $\partial_2 u_1$ is more intermittent than
the longitudinal one $\partial_1 u_1$, see e.g.\ \cite{ker85}
-- in our simulations we have flatnesses of
$F_{\partial_1 u_1 } = 4.9$,
$F_{\partial_2 u_1 } = 7.0$,
$F_{\partial_1 u_2 -\partial_2 u_1} = 7.0$ -- 
it is not trivial that this difference, probing the VSR,
is carried on into the
ISR. In ref.\ \cite{bor97} the different degrees of intermittency were
associated with different types of structures:
Longitudinal fluctuations with strain like structures,
transversal fluctuations with vorticity like structures -- which both
is in keeping with
the definitions of strain and vorticity, respectively.

Let us discuss our results on $\dx_p^{L,T}$ in figure \ref{fig_zetasl}
in more detail. The values for $\dx_p^L$ are well described by the SL model
fit eq.\ (\ref{sl})
with the SL values $C_0 = 2$, $\beta = (2/3)^{1/3}\approx 0.874$
as found for many other
isotropic, even low $\rel$ number, experimental or numerical flows
\cite{fri95,ben93b,ben95}. For the 
physical interpretation of the parameters in the SL model we refer to refs.\
\cite{she94,she95}. From a phenomenological point of view, one could
consider eq.\ (\ref{sl}) simply as a two parameter fit of the $\xi_p$'s.
The two SL parameters for the {\it transversal} scaling exponents can 
be viewed as a  simple way to quantify
the degree of intermittency.

\begin{figure}[htb]
\setlength{\unitlength}{1.0cm}
\begin{picture}(9,9)
\put(0.5,0.5)
{\psfig{figure=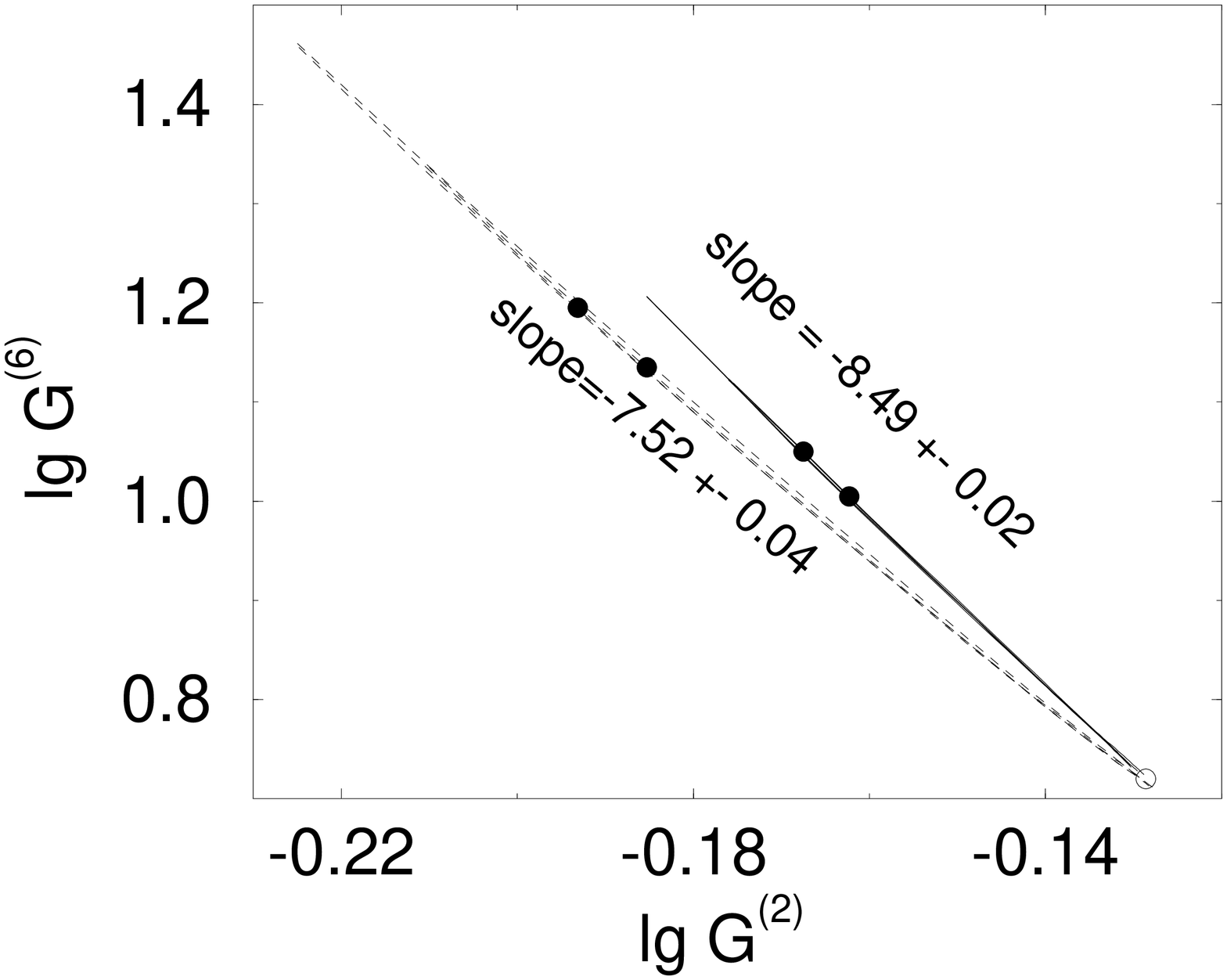,width=10cm,angle=-90}}
\end{picture}
\caption[]{
GESS type plot 
$G^{(6)}(r)$ vs $G^{(2)}(r)$ for both the longitudinal (solid) and the
transversal (dashed)
 G-structure functions for the isotropic 
 simulations.
The slopes of these curves are $\rho^L_{6,2} = -8.49 \pm 0.02$ and
$\rho_{6,2}^T = -7.52 \pm 0.04$, respectively.
The errors result from  a linear regression of every single curve,
from weighted averaging of the results for different space
directions, and different $\rel$.
 The hardly distinguishable
lines within the  two bunches of curves are the results for different
directions and different Reynolds numbers $\rel =110 $ and $\rel =70$.
The good agreement within the bunches means good isotropy and
independence of the scaling exponents from $\rel$.
Note that in this type of plot the far VSR collapses
into the upper left point of these curves.
The filled bullets  refer to $r=10\eta$ for $\rel = 110$ (left one)
and $\rel = 70$ (right one). The open bullet  refers to the outer
length scale $r= L$.  
}
\label{fig_gess}
\end{figure}

\begin{figure}[htb]
\setlength{\unitlength}{1.0cm}
\begin{picture}(9,9)
\put(0.5,0.5)
{\psfig{figure=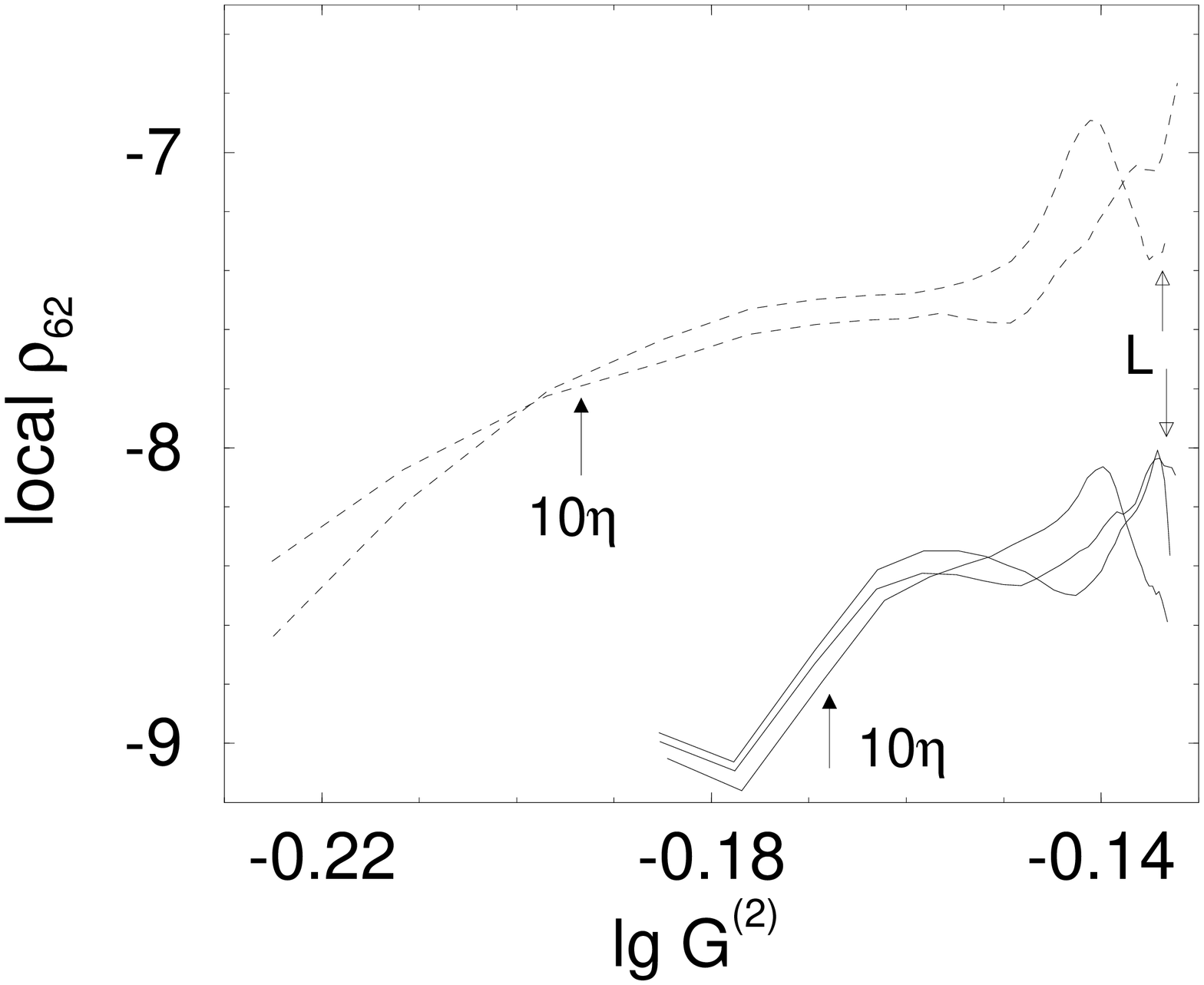,width=10cm,angle=-90}}
\end{picture}
\caption[]{
Local slopes of figure \ref{fig_gess} for the $\rel = 110$ simulation.
The two dashed lines are for the two different space direction
for the transversal structure functions, the three solid lines
are for the three space directions of the longitudinal structure functions.
If we calculate the average (for scales up to $r=2.0 \approx 100\eta$)
we obtain 
$\rho_{6,2}^L = -8.36 \pm 0.14$
and $\rho_{6,2}^T = -7.36 \pm 0.26$. The arrows 
refer to $10\eta$ and $L$, respectively. 
}
\label{fig_gess_local}
\end{figure}

We now suggest a method to replace this one two-parameter-fit by two
one-parameter fits.
To do so, we employ generalized extended self similarity 
(GESS, \cite{ben95,ben96b}) and plot $G_p$ vs $G_q$, see figure 
\ref{fig_gess}. The slope $\rho_{p,q}$ of such a plot is by definition
(\ref{gfct})
\be
\rho_{p,q} = {\xi_p -  p/3 \over \xi_q - q /3}.
\label{eq_rho}
\ee
For fixed $p,q$,
$\rho_{p,q}^L$ and
$\rho_{p,q}^T$
are {\it significantly different},
see table \ref{tab_rho}. We checked this result very
carefully. The small error bars in $\rho_{p,q}$ result 
from  linear regressions in  GESS plots as in
figure \ref{fig_gess} and in addition from 
an averaging over the different space
directions. We also checked this for smaller $\rel = 70$;
the deviations in comparison to the
results given in table \ref{tab_rho} are never larger than $0.5\%$.
Moreover, to make
sure that our numbers  are well converged, we also  averaged over
only $30$,
$60$, and 
$90$
large eddy
turnovers rather than 140; still, the result is the same; the deviations
are smaller than the error bars.

To quantify the
quality to which GESS holds we checked the relation
$\rho_{p,s} = \rho_{p,q} \rho_{q,s}$, implied by GESS, for various
$p,q$.
Table \ref{tab_rho} allows the reader to do so. Neither for the
$\rho_{p,q}^L$ nor for the 
$\rho_{p,q}^T$ did we find a single example where there were deviations
larger than the error bars. E.g., 
$\rho_{2,4}^L \cdot 
\rho_{4,6}^L = 0.1177 \pm 0.0005 $ which equals 
$\rho_{2,6}^L$ within the error bars. 

The error bars up to now stem from {\it statistics}. One would like to be
able to judge the size of the {\it systematic} errors.
Therefore, in figure \ref{fig_gess_local}
we display the {\it local slope} of the curves in fig.\ \ref{fig_gess}.
Both the longitudinal and the transversal local slope
 slightly increase  (modulus-wise)
with increasing scale (i.e., from
left to right), which shows the limitations of
 the above statement that GESS is
fulfilled with remarkable quality. The error bar calculated
from averaging the local slope is much bigger than above statistical
error. From averaging up to the scale $r=2.0 \approx 100\eta$ we obtain
$\rho_{6,2}^L = -8.36 \pm 0.14$
and $\rho_{6,2}^T = -7.36 \pm 0.26$. The numbers
for the $\rho_{p,q}^{L,T}$ from table \ref{tab_rho}
are within the (now about ten times larger) error bars.
These systematic errors 
are one order of magnitude bigger than
the purely
statistical ones in table  \ref{tab_rho}.
We summarize the values of $\rho_{p,q}^{L,T}$ (and their error bars)
determined in this
way in table \ref{tab_rho_sys}. 
Note, however, that the
deviations between
$\rho_{p,q}^L$ and 
$\rho_{p,q}^T$ and correspondingly also between the resulting $\beta$'s
(see below) 
are still statistically significant.

 \begin{table}[t]
 \begin{center}
 \begin{tabular}{|c|c|c|c|c|}
 \hline
          $p,q$
       &  $\rho_{p,q}^L$
       & $\beta^L$
       & $\rho_{p,q}^{T}$
       & $\beta^T$
 \\
\hline
$ 2,4 $
& $ -0.5093 \pm 0.0004 $
& $ 0.973 \pm 0.001 $
& $ -0.5440 \pm 0.0011 $
& $ 0.877 \pm 0.003 $
\\
$ 2,5 $
& $ -0.2073 \pm 0.0002 $
& $ 0.964 \pm 0.001 $
& $ -0.2272 \pm 0.0006 $
& $ 0.875 \pm 0.002 $
\\
$ 2,6 $
& $ -0.1176 \pm 0.0002 $
& $ 0.957 \pm 0.001 $
& $ -0.1318 \pm 0.0004 $
& $ 0.873 \pm 0.002 $
\\
$ 2,7 $
& $ -0.0774 \pm 0.0001 $
& $ 0.952 \pm 0.001 $
& $ -0.0884 \pm 0.0003 $
& $ 0.872 \pm 0.002 $
\\
$ 2,8 $
& $ -0.0555 \pm 0.0001 $
& $ 0.947 \pm 0.001 $
& $ -0.0645 \pm 0.0003 $
& $ 0.871 \pm 0.003 $
\\
$ 4,5 $
& $ 0.4072 \pm 0.0001 $
& $ 0.947 \pm 0.001 $
& $ 0.4177 \pm 0.0003 $
& $ 0.870 \pm 0.002 $
\\
$ 4,6 $
& $ 0.2311 \pm 0.0001 $
& $ 0.941 \pm 0.001 $
& $ 0.2423 \pm 0.0004 $
& $ 0.869 \pm 0.003 $
\\
$ 4,7 $
& $ 0.1521 \pm 0.0002 $
& $ 0.937 \pm 0.001 $
& $ 0.1625 \pm 0.0005 $
& $ 0.868 \pm 0.003 $
\\
$ 4,8 $
& $ 0.1092 \pm 0.0003 $
& $ 0.933 \pm 0.002 $
& $ 0.1186 \pm 0.0005 $
& $ 0.867 \pm 0.003 $
\\
$ 5,6 $
& $ 0.5675 \pm 0.0002 $
& $ 0.935 \pm 0.001 $
& $ 0.5800 \pm 0.0005 $
& $ 0.867 \pm 0.003 $
\\
$ 5,7 $
& $ 0.3736 \pm 0.0004 $
& $ 0.931 \pm 0.002 $
& $ 0.3890 \pm 0.0008 $
& $ 0.867 \pm 0.003 $
\\
$ 5,8 $
& $ 0.2682 \pm 0.0006 $
& $ 0.928 \pm 0.002 $
& $ 0.2840 \pm 0.0010 $
& $ 0.866 \pm 0.004 $
\\
$ 6,7 $
& $ 0.6583 \pm 0.0005 $
& $ 0.927 \pm 0.002 $
& $ 0.6708 \pm 0.0008 $
& $ 0.865 \pm 0.004 $
\\
$ 6,8 $
& $ 0.4725 \pm 0.0008 $
& $ 0.924 \pm 0.003 $
& $ 0.4901 \pm 0.0014 $
& $ 0.863 \pm 0.005 $
\\
$ 7,8 $
& $ 0.7177 \pm 0.0007 $
& $ 0.922 \pm 0.003 $
& $ 0.7308 \pm 0.0010 $
& $ 0.860 \pm 0.005 $
\\
 \hline
 \end{tabular}
 \end{center}
\caption{{\it
$\rho_{p,q}^{L,T}$ for various pairs $p,q$ from GESS type plots as in
figure 6.
By definition, $\rho_{p,q} = \rho_{q,p}^{-1}$.
The errors are purely statistical ones. 
In the third and fifth column, we
give the $\beta$'s resulting from 
eq.\ (12).
Note that GESS implies $\rho_{p,s} = \rho_{p,q} \rho_{q,s}$.
This relation can be used to check the quality of GESS.
}}
\label{tab_rho}
 \end{table}

 \begin{table}[t]
 \begin{center}
 \begin{tabular}{|c|c|c|c|c|}
 \hline
          $p,q$
       &  $\rho_{p,q}^L$
       & $\beta^L$
       & $\rho_{p,q}^{T}$
       & $\beta^T$
 \\
\hline
$ 2,4 $
& $ -0.514 \pm 0.005 $
& $ 0.958 \pm 0.016 $
& $ -0.556 \pm 0.012 $
& $ 0.846 \pm 0.031 $
\\
$ 2,5 $
& $ -0.210 \pm 0.003 $
& $ 0.950 \pm 0.014 $
& $ -0.234 \pm 0.007 $
& $ 0.848 \pm 0.027 $
\\
$ 2,6 $
& $ -0.120 \pm 0.002 $
& $ 0.944 \pm 0.012 $
& $ -0.136 \pm 0.005 $
& $ 0.849 \pm 0.025 $
\\
$ 2,7 $
& $ -0.079 \pm 0.002 $
& $ 0.940 \pm 0.012 $
& $ -0.092 \pm 0.003 $
& $ 0.851 \pm 0.023 $
\\
$ 2,8 $
& $ -0.057 \pm 0.001 $
& $ 0.936 \pm 0.011 $
& $ -0.067 \pm 0.003 $
& $ 0.852 \pm 0.021 $
\\
$ 4,5 $
& $ 0.409 \pm 0.001 $
& $ 0.936 \pm 0.011 $
& $ 0.420 \pm 0.003 $
& $ 0.851 \pm 0.021 $
\\
$ 4,6 $
& $ 0.233 \pm 0.002 $
& $ 0.931 \pm 0.010 $
& $ 0.245 \pm 0.003 $
& $ 0.853 \pm 0.020 $
\\
$ 4,7 $
& $ 0.153 \pm 0.002 $
& $ 0.927 \pm 0.010 $
& $ 0.165 \pm 0.003 $
& $ 0.855 \pm 0.018 $
\\
$ 4,8 $
& $ 0.110 \pm 0.001 $
& $ 0.924 \pm 0.011 $
& $ 0.120 \pm 0.003 $
& $ 0.856 \pm 0.018 $
\\
$ 5,6 $
& $ 0.569 \pm 0.002 $
& $ 0.926 \pm 0.010 $
& $ 0.582 \pm 0.003 $
& $ 0.855 \pm 0.018 $
\\
$ 5,7 $
& $ 0.376 \pm 0.002 $
& $ 0.923 \pm 0.010 $
& $ 0.391 \pm 0.004 $
& $ 0.857 \pm 0.018 $
\\
$ 5,8 $
& $ 0.270 \pm 0.003 $
& $ 0.920 \pm 0.011 $
& $ 0.286 \pm 0.004 $
& $ 0.858 \pm 0.018 $
\\
$ 6,7 $
& $ 0.660 \pm 0.002 $
& $ 0.919 \pm 0.011 $
& $ 0.672 \pm 0.003 $
& $ 0.858 \pm 0.018 $
\\
$ 6,8 $
& $ 0.475 \pm 0.004 $
& $ 0.917 \pm 0.012 $
& $ 0.491 \pm 0.005 $
& $ 0.859 \pm 0.018 $
\\
$ 7,8 $
& $ 0.719 \pm 0.003 $
& $ 0.914 \pm 0.014 $
& $ 0.731 \pm 0.004 $
& $ 0.860 \pm 0.019 $
\\
 \hline
 \end{tabular}
 \end{center}
\caption{{\it
$\rho_{p,q}^{L,T}$ for various pairs $p,q$
determined from the local slope of the GESS type plots, cf.\ figure
7.
The errors are the systematic ones, stemming from the local
slope not being constant. 
In the third and fifth column, we
again give the $\beta$'s and their errors resulting from 
eq.\ (12). 
Averaging the $\beta$'s determined this way gives 
$\beta^L = 0.930 $ and
$\beta^T = 0.855 $.
}}
\label{tab_rho_sys}
 \end{table}

Within the SL model, the $\rho_{p,q}$'s only depend on $\beta$,
not on any other parameter,
\be
\rho_{p,q}^{SL} = {
(1-\beta^{p}) - (p/3)(1-\beta^3 ) \over
(1-\beta^{q}) - (q/3)(1-\beta^3 )}. 
\label{rho_sl}
\ee
For each $\rho_{p,q}$ we calculate $\beta_{p,q}$, resulting from
equation (\ref{rho_sl}), and its error, see table \ref{tab_rho}
and table \ref{tab_rho_sys}.
If the SL model were exact, $\beta$ should not depend on $p$ and $q$.

\begin{figure}[htb]
\setlength{\unitlength}{1.0cm}
\begin{picture}(9,9)
\put(0.5,0.5)
{\psfig{figure=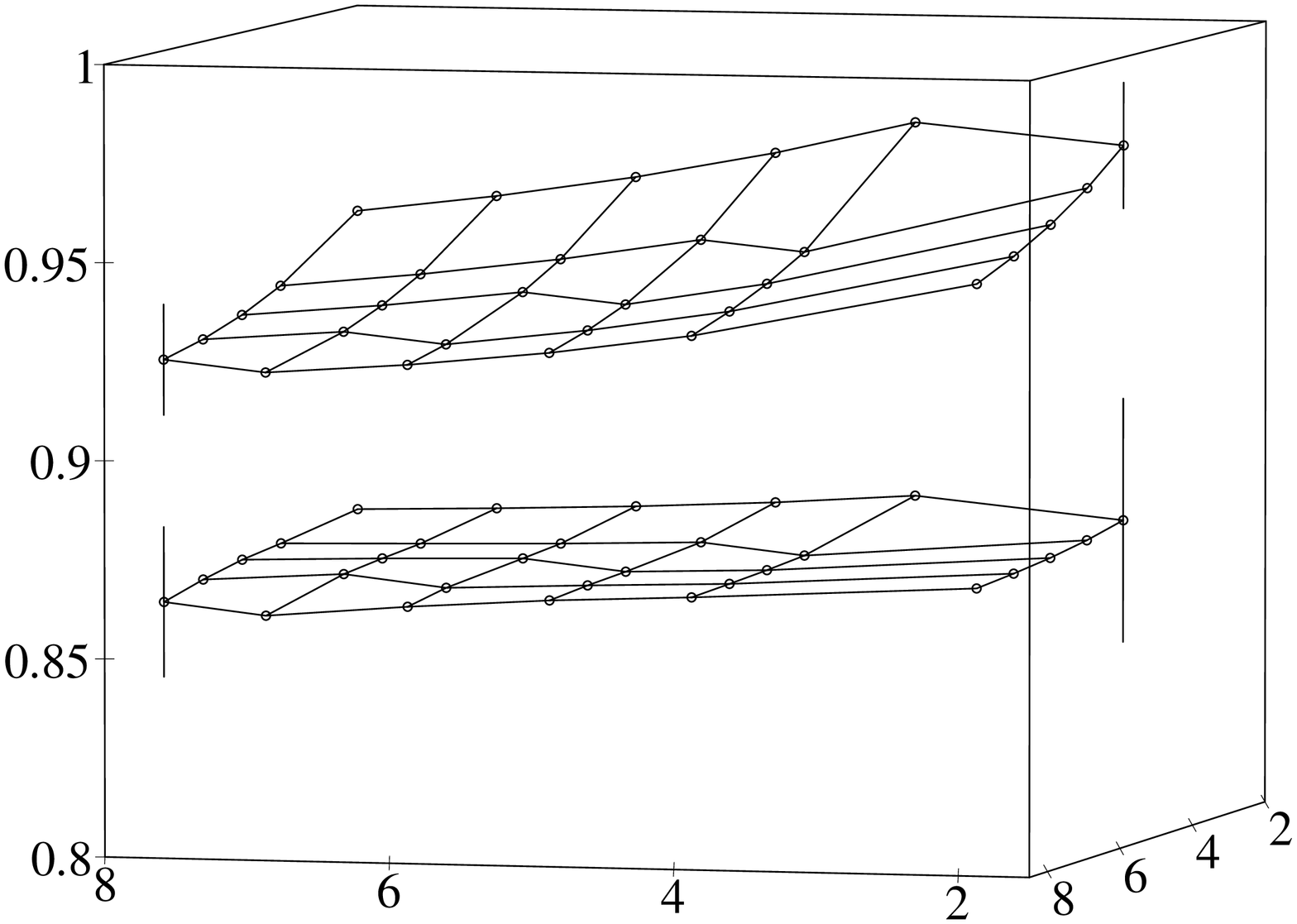,width=10cm,angle=-90}}
\put(4.4,1.0){\large$p$}
\put(8.7,1.3){\large$q$}
\put(0.5,4.0){\large$\beta$}
\end{picture}
\caption[]{
The longitudinal and transversal hierarchy parameters
$\beta_{p,q}^{L}$ (upper) and 
$\beta_{p,q}^{T}$ (lower), respectively. The data are 
taken from table \ref{tab_rho}, i.e., for the ``I'' simulation with
$\rel = 110$, the averaging time is 140 large eddy turnovers.
To get an idea of the size of the error, two error bars are drawn,
representing the much larger {\it systematic} errors rather than
the statistical ones. 
First, we observe that $\beta^L$ and $\beta^T$ are clearly different.
Second, a slight dependence of $\beta^{L,T}$ on $p,q$ is seen.
}
\label{fig_beta}
\end{figure}

In figure \ref{fig_beta}
we offer a 3D plot of $\beta_{p,q}^{L,T}$, together with
the error bars resulting from the (larger)
{\it systematic} errors of $\rho$, cf.\ table \ref{tab_rho_sys}. 
From figure \ref{fig_beta} the difference between $\beta^L$ and $\beta^T$
seems to be significant. This result is at variance with Camussi and Benzi's
\cite{cam97} who obtained that the difference of both 
$\beta^L$ and $\beta^T$ to the SL value $\beta = (2/3)^{1/3} \approx 0.874$
is at most $1.2\%=0.010$.

Another feature of figure \ref{fig_beta} is that $\beta_{p,q}^L$ shows a small
trend towards smaller values for larger $p,q$ which is not expected within the
SL model.
If we 
average over all $\beta_{p,q}$ nevertheless, we obtain
$\beta^L = 0.947 $ and $\beta^T = 0.870 $.

\begin{figure}[htb]
\setlength{\unitlength}{1.0cm}
\begin{picture}(9,9)
\put(0.5,0.5)
{\psfig{figure=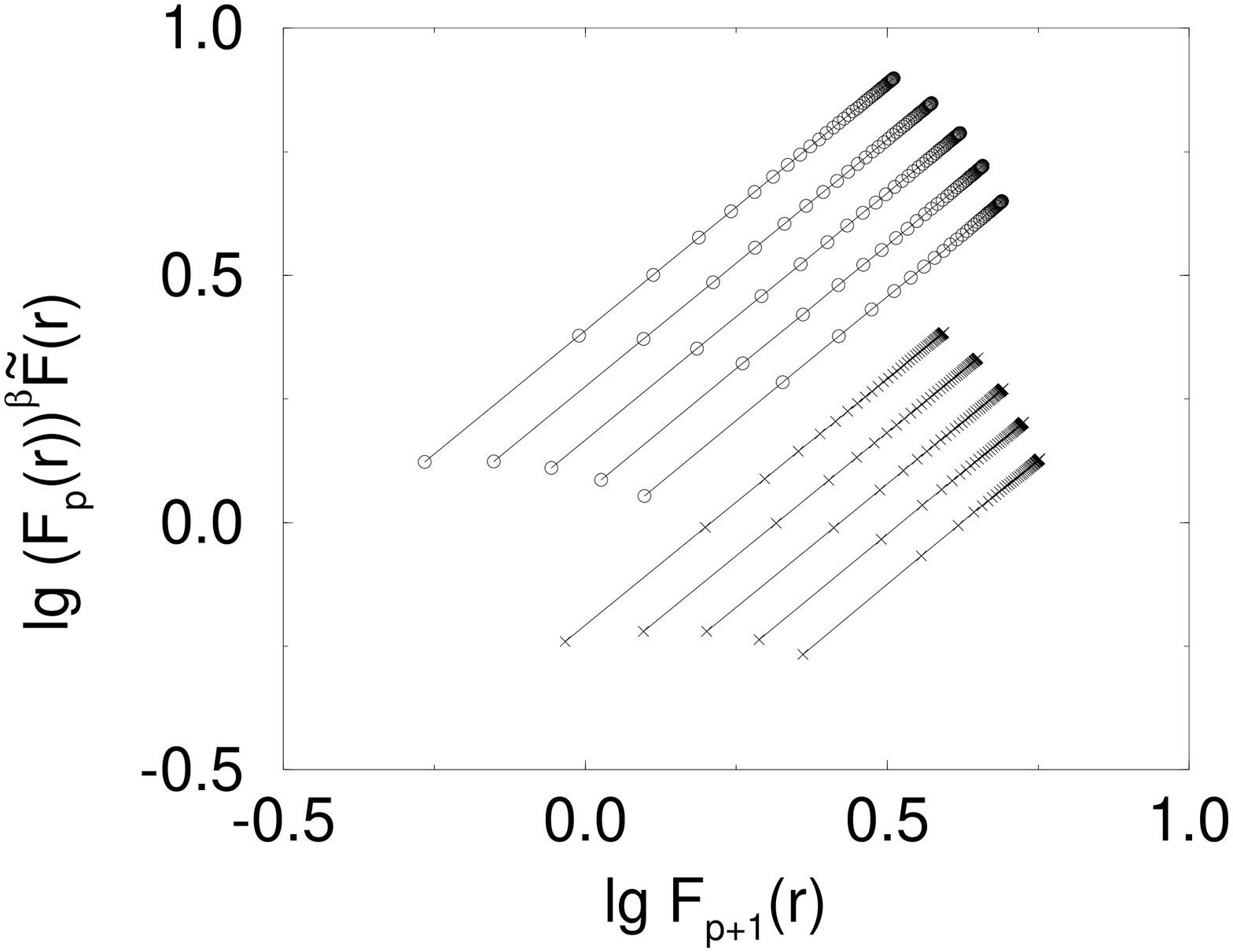,width=10cm,angle=-90}}
\end{picture}
\caption[]{
Log-Log plot of
$F_{p+1} (r)$ vs
$(F_p(r))^\beta \tilde F(r)$
for various $p$ for the longitudinal and transversal structure functions,
$\beta^L = 0.947$, $\beta^T = 0.870$. The lines
are arbitrarily shifted 
in order for the slopes to be visible. The upper 5 lines are for
the longitudinal structure functions, $p=3$ (upper)
to $p=7$ (lower), the lower 5 lines
are for 
the transversal structure functions, $p=3$ (upper) to $p=7$ (lower). 
}
\label{fig_ruiz}
\end{figure}

 \begin{table}[htp]
 \begin{center}
 \begin{tabular}{|c|c|c|c|c|c|c|}
 \hline
          $$
       &  $\beta^L = 0.947$
       & $$
       & $\beta^T=0.870$
       & $$
       & $\beta^L=(2/3)^{1/3}$
       & $$
 \\
          $p$
       &  slope $L$
       & $B_p^L$
       &  slope $T$
       & $B_p^T$
       & slope $L$
       & $B_p^L$
 \\
\hline
$ 3 $
& $ 0.9984 \pm 0.0001 $ & $ 0.975 $
& $ 0.9988 \pm 0.0002 $ & $ 0.981 $
& $ 0.9988 \pm 0.0001 $ & $ 0.981 $
\\
$ 4 $
& $ 0.9995 \pm 0.0001 $ & $ 0.944 $
& $ 0.9993 \pm 0.0004 $ & $ 0.961 $
& $ 0.9936 \pm 0.0001 $ & $ 0.969 $
\\
$ 5 $
& $ 1.0020 \pm 0.0004 $ & $ 0.925 $
& $ 1.0000 \pm 0.0005 $ & $ 0.951 $
& $ 0.9895 \pm 0.0002 $ & $ 0.967 $
\\
$ 6 $
& $ 1.0050 \pm 0.0007 $ & $ 0.910 $
& $ 1.0020 \pm 0.0007 $ & $ 0.945 $
& $ 0.9854 \pm 0.0004 $ & $ 0.970 $
\\
$ 7 $
& $ 1.0080 \pm 0.0010 $ & $ 0.899 $
& $ 1.0040 \pm 0.0016 $ & $ 0.940 $
& $ 0.9810 \pm 0.0007 $ & $ 0.977 $
\\
\hline
 \end{tabular}
 \end{center}
\caption{{\it
Slopes of
$\lg F_{p+1} (r)$ vs
$\lg ((F_p(r))^\beta \tilde F(r))$, varying $r$. 
According to eq.\ (13), the slopes
should be 1, which is pretty well fulfilled for $\beta^L = 0.947$ and
$\beta^T = 0.870$. For comparision, we also give the slopes if
the SL-value 
$\beta^L = (2/3)^{1/3}$ is used for the longitudinal structure function.
The deviations of the slope to 1 are larger.
The constants $B_p$ in (13) show a slight $p$-dependence. 
}}
\label{tab_ruiz}
 \end{table}

Knowing $\beta$, 
there is only the parameter
$C_0$ left in eq.\ (\ref{sl}). 
If we take the above mean values $\beta^L = 0.947$ and
$\beta^T = 0.870$, we obtain as 
best fits to the $\xi_p$ data in figure \ref{fig_zetasl} 
$C_0^L = 9.3 $ (the $\chi^2$ of the fit is
$\chi^2 = 10$) and 
$C_0^T = 3.7$ (with $\chi^2 = 1$),
excellently describing the numerical data.
We do not ascribe any
physical meaning to the parameter values obtained in
our fit.
Note that for our $\xi^L$ data this fit is superior to the SL
model with the original parameter values $\beta =(2/3)^{1/3}$, $C_0 =2$.
If we  choose the SL-value $\beta^L = \beta^T = (2/3)^{1/3}$ we obtain
$C_0^L = 1.97$ with $\chi^2 = 10^{3}$
and $C_0^T = 3.9$ with $\chi^2 = 0.8$.
(The $\chi^2$-values for $C_0^L$ are larger than those for $C_0^T$
as the errors of the $\delta \xi_p^L$ are smaller than those
of $\delta \xi_p^T$.)

We now directly check the hierarchies of the structure functions
\cite{she94,rui95,ben96b}. 
 From eq.\ (\ref{d_hier}) it is easy to derive \cite{rui95,ben96b}
 \be
 F_{p+1} (r) = B_p \left( F_p(r)\right)^\beta \tilde F(r)
 \label{r1}
 \ee
with
\be
F_{p+1} = { D_{p+1} (r) \over D_p (r) }
\label{r2}
\ee
and
\be
\tilde F(r) = \left( {D_6(r) \over (D_3^*(r))^{1+\beta^3 }}
\right)^{(1-\beta)/[3(1-\beta^3 )]}.
\label{r3}
\ee
With the mean $\beta$'s obtained above, we plot
$F_{p+1} (r)$ vs
$(F_p(r))^\beta \tilde F(r)$. If eq.\ (\ref{r1}) and
equivalently eq.\ (\ref{d_hier}) hold, the slope should be 1. Indeed, the slope
is very close to 1, see figure \ref{fig_ruiz} and table \ref{tab_ruiz},
which gives further support for $\beta^L$ and $\beta^T$ being different. 
The best agreement is found for $p=4$ to $p=6$, see table \ref{tab_ruiz}.
The reason is that the mean $\beta$'s best agree with $\beta_{p,q}$ if 
$p,q$ are around 4 -- 6,
see table \ref{tab_rho}. For the other $p$ one could improve
the fit by using the corresponding $\beta_{p,q}$; however, note that the
sixth order structure function always enters via $\tilde F(r)$, cf.\ eq.\
(\ref{r3}).

Moreover, we find a $p$ dependence of the prefactor $B_p$ in eq.\
(\ref{r1}).
Therefore, determining $\beta$ from eq.\ (\ref{r1}) by plotting
$\lg F_{p+1}(r)$ vs $\lg F_p(r)$ for fixed $r$
as a function of $p$ as done in refs.\
\cite{ben96c,cam96} does not seem to be possible here.

\begin{figure}[htb]
\setlength{\unitlength}{1.0cm}
\begin{picture}(9,9)
\put(0.5,0.5)
{\psfig{figure=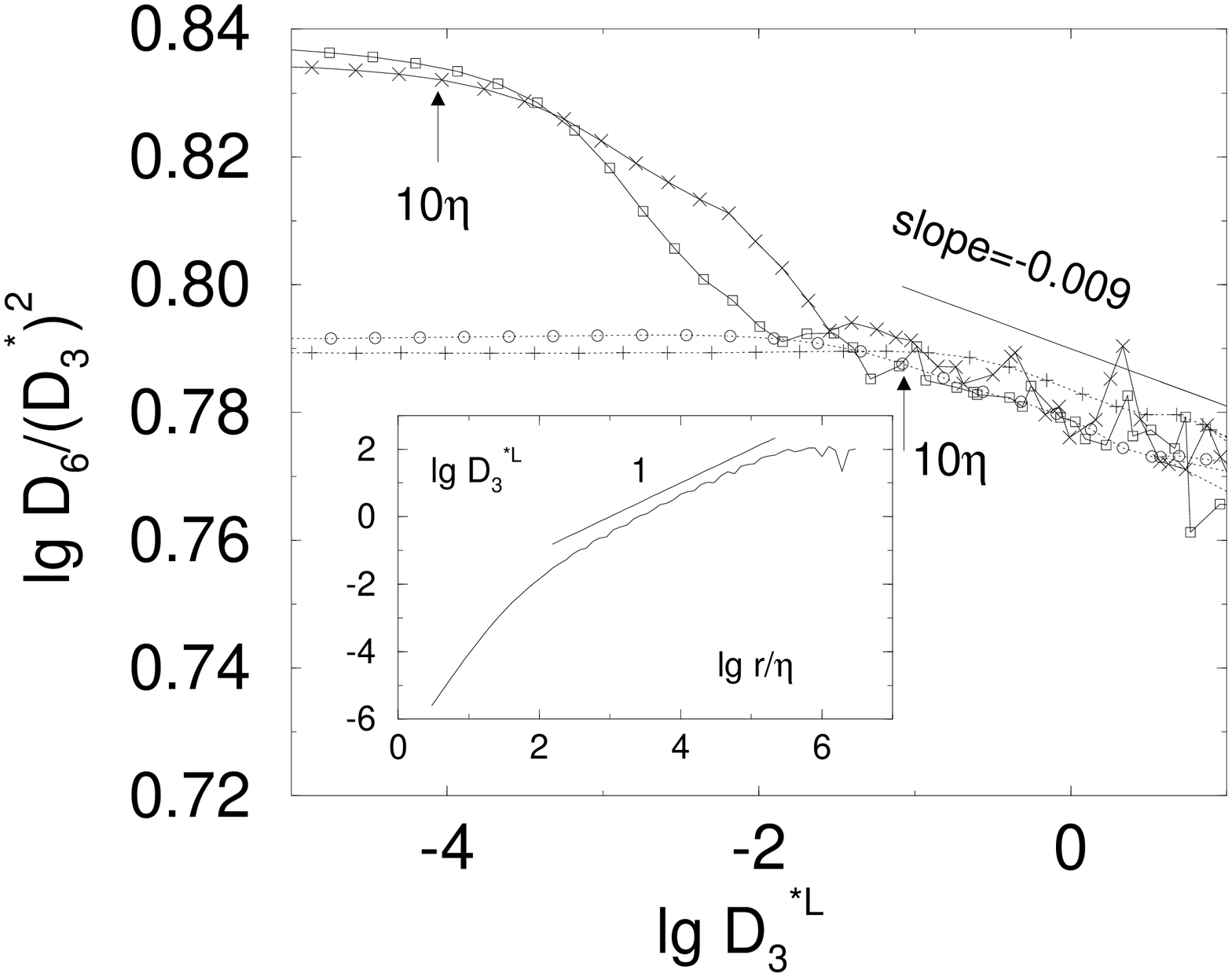,width=10cm,angle=-90}}
\end{picture}
\caption[]{
Compensated ESS plot for the sixth order
structure functions for the REWA calculation with 50 modes per level.
The longitudinal and the transversal
structure functions show the {\it same} slope $\dx_p = -0.009$.
Squares are for $\rel = 1.4 \cdot 10^5$, longitudinal;
crosses for $\rel = 1.4 \cdot 10^5$, transversal;
circles for $\rel = 8\cdot 10^2$, longitudinal;
pluses for $\rel = 8\cdot 10^2$, transversal.
The two arrows indicate the scale $10\eta$ for the simulation with
the higher (left arrow) and the lower (right arrow) Reynolds number,
respectively.  
The inset shows the third order structure function $D_3^{*L} (r)$
(for $\rel = 1.4 \cdot 10^5$)
in order to give an idea to what scale $r/\eta$ the data in the figure
correspond. 
}
\label{fig_rewa_ess}
\end{figure}

\section{Scaling relations within REWA}
Very large $\rel$ in numerical turbulent flow 
can be achieved in the reduced wave vector set
approximation  (REWA) of the Navier-Stokes equation
\cite{egg91a,gnlo92b,gnlo94a,gro95b,uhl97}.
REWA uses 
a reduced, geometrically scaling
subset of wavevectors on which the Navier-Stokes equation is solved.
Here we choose a basic set of 50 modes per level.
Very high
Taylor-Reynolds numbers up to $Re_\lambda = 7\cdot 10^4$
\cite{gnlo94a,gro95b} can be achieved, however, flow structures are
underrepresented \cite{gro96} and the intermittency corrections are
strongly underestimated \cite{gnlo94a,uhl97}.

We redid ESS types plots for REWA for
 $\rel = 8 \cdot 10^2$
and for
$\rel= 1.4\cdot 10^5$ 
for both
the longitudinal and the transversal sixth order structure functions, see
figure \ref{fig_rewa_ess}.  There is no detectable difference 
between the longitudinal and transversal scaling exponents.
The absolute value $\dx_6^L \sim \dx_6^T \sim -0.009$
 is much smaller
 (modulo wise) than the experimental or above numerical 
value $\dx_6^L \sim -0.21$, as extensively analyzed and discussed in
the previous work on REWA
\cite{egg91a,gnlo92b,gnlo94a,gro95b,uhl97}.
Note however that the relative
error for the $\dx$'s 
 is much larger than 
in the full numerical simulations -- 
we cannot exclude different degrees of intermittency for
$D_p^L$ and $D_p^T$ within REWA.

We do not know whether our results on REWA indicate that the
differences between the scaling of $D^T_p$ vs $D_3^{*T}$
and $D^L_p$ vs $D_3^{*L}$ observed in the
above full numerical simulations for small $\rel$
are finite $\rel$ effects or whether they are  artefacts of the
REWA thinning of large wave vectors \cite{egg91a,gnlo94a,gro96},
connected to the suppression of small scale structures. Such structures
are
associated with the different scaling of longitudinal and transversal
structure functions in ref.\ \cite{bor97}.
As within REWA no $\rel$ dependence of the $\dx^L$ and $\dx^T$
 is observed, see figure \ref{fig_rewa_ess},
 we favor the second interpretation.

We also tried GESS type scaling within REWA. No statistically significant
deviations  between the
$\rho_{p,q}^L$ and the 
$\rho_{p,q}^T$ were found.

\section{Summary, conclusions, and outlook}
To summarize, we offer strong evidence
that the transversal velocity fluctuations 
show stronger intermittency than the longitudinal ones.
Our numerical
values for the longitudinal and the transversal
scaling exponents $\xi_p^{L}$ and $\xi_p^T$
for forced stationary
turbulence agree very well with those of
Boratav and Pelz for decaying turbulence \cite{bor97}, see table
\ref{tab_exp}. 
This finding is independent of $\rel$, at least for the relatively
low $\rel$ we examined.
For an anisotropic flow we essentially
obtained the same scaling exponents. 
Only for the
REWA calculations which underrepresent the small scale
structures of the flow
we do {\it not} find a statistically significant
deviation between $\dx_p^L$ and $\dx_p^T$, however,
the relative error is much bigger than for the full simulations.

We reiterate that $\zeta_2^L = \zeta_2^T$ because of relation (\ref{eq4});
a generalization of this equality to higher order moments $p>2$ is wrong. 

GESS is fulfilled with satisfactory precision for both longitudinal
and transversal structure functions. The
GESS scaling exponents
$\rho^{L}_{p,q}$
and $\rho^{T}_{p,q}$
are different.
This result is the more remarkable, as those of the longitudinal
velocity structure function agree with those of
an active \cite{ben94b} or passive scalar \cite{ben96b}
or even with those calculated for
the magnetic field in MHD \cite{gra94}, see table 1 of ref.\ \cite{ben96b}.

The She-Leveque hierarchy parameters
$\beta^{L,T}$ following from the $\rho^{L,T}$-exponents  
consequently also differ. Both show a weak dependence on $p,q$ which is
not expected within the She-Leveque model.

In the whole analysis we took great care of systematic and
statistical errors to get significant statements. 

To conclude, there seem to exist  independently scaling velocity fields
$v^L(r)$ and $v^T(r)$, i.e., the Navier-Stokes dynamics seems to make use of
this degree of freedom being allowed by symmetry. It is very likely
that the two different scaling
velocities fields will also be reflected in the flow geometry.

A more complete discription 
of the statistics of the velocity field was recently suggested by
L'vov, Podivilov, and Procaccia \cite{lvov97}.
These authors 
point out that because of rotational
symmetry only the SO(3) irreducible amplitudes of the velocity
structure tensors should obey clean scaling. In ref.\ \cite{us}
we analyse the scaling properties of these amplitudes numerically and
show (for fourth order moments) how they are connected to the
longitudinal and transversal structure functions.

Presently, neither the longitudinal nor the transversal 
scaling exponents,
nor the scaling exponents of the irreducible SO(3) invariants of the
velocity correlations \cite{lvov97} 
 can be calculated analytically from the Navier-Stokes
dynamics. Many phenomenological models based on various views on how
intermittency develops are able to fit the {\it longitudinal} intermittency
corrections.  It should be possible to derive also the {\it transversal}
intermittency corrections in the framework of the thinking these models are
based on
in order to check their consistency.
 The ultimate goal, however, must be to derive both longitudinal and
transversal scaling exponents from the Navier-Stokes equation.

\vspace{1cm}

\noindent
{\bf Acknowledgments:}
We thank
Luca Biferale and O.\ Boratav for careful reading of the manuscript and 
Rolf Nicodemus for drawing figure \ref{fig_beta}.
Support for this work by
the Deutsche Forschungsgemeinschaft (DFG) under grant SBF185-D3 and by
the German-Israel Foundation (GIF) is acknowledged.
The HLRZ J\"ulich supplied us with computer time. 
D.\ L.\ acknowledges the hospitality of Leo Kadanoff and the
University of Chicago, where part of the work was done, and support by
MRSEC.

\vspace{0.5cm}

\noindent
 e-mail addresses:\\
grossmann$\_$s@stat.physik.uni-marburg.de\\ 
lohse@stat.physik.uni-marburg.de\\
reeh@mailer.uni-marburg.de



\end{document}